\documentclass[a4paper,11pt]{article}
\usepackage{pos}

\title{Quasi-free (p,2p) reactions in inverse kinematics for studying the fission yield dependence on temperature}

\author[1]{A. Gra\~{n}a-Gonz\'{a}lez} 
\author[1]{J. L. Rodr\'{i}guez-S\'{a}nchez}
\author[1]{J. Benlliure}
\author[1]{G. Garc\'{i}a-Jim\'{e}nez}
\author[1]{H. Alvarez-Pol}
\author[1]{D. Cortina-Gil}
\author[2]{L. Atar}
\author[3]{L. Audouin}
\author[4]{G. Authelet}
\author[5]{A. Besteiro}
\author[6]{G. Blanchon}
\author[7]{K. Boretzky}
\author[1]{P. Cabanelas}
\author[5]{E. Casarejos}
\author[8]{J. Cederkall}
\author[6]{A. Chatillon}
\author[4]{A. Corsi}
\author[1]{M. Feijoo}
\author[9]{D. Galaviz}
\author[10]{I. Gasparic}
\author[11]{R. Gernh{\"a}user}
\author[7]{M. Heil}
\author[12]{A. Heinz}
\author[12]{M. Holl}
\author[11]{T. Jenegger}
\author[2]{L. Ji}
\author[12]{H.T. Johansson}
\author[7]{O.A. Kiselev}
\author[11]{P. Klenze}
\author[8]{A. Knyazev}
\author[7]{D. K{\"o}rper}
\author[2]{T. Kr{\"o}ll}
\author[10]{I. Lihtar}
\author[7]{Y. A. Litvinov}
\author[7]{B. L{\"o}her}
\author[6]{P. Morfouace}
\author[13,14]{D. M{\"u}cher}
\author[15]{S. Murillo Morales}
\author[2]{A. Obertelli}
\author[7]{V. Panin}
\author[16]{J. Park}
\author[15]{S. Paschalis}
\author[17]{A. Perea}
\author[15]{M. Petri}
\author[18]{S. Pirrone}
\author[11]{L. Ponnath}
\author[6]{A. Revel}
\author[2]{H.-B. Rhee}
\author[15]{L. Rose}
\author[2]{D.M. Rossi}
\author[18]{P. Russotto}
\author[7]{H. Simon}
\author[15]{A. Stott}
\author[2]{Y. Sun}
\author[2]{C. S{\"u}rder}
\author[6]{J. Taieb}
\author[15]{R. Taniuchi}
\author[17]{O. Tengblad}
\author[2,7]{H.T. T\"ornqvist}
\author[2]{S. Velardita}
\author[19]{J. Vesic}
\author[7]{B. Voss}
\author[]{the R³B collaboration.}

\affiliation[1]{IGFAE, University of Santiago de Compostela, E-15782 Santiago de Compostela, Spain}

\affiliation[2]{Technische Universit{\"a}t Darmstadt, Fachbereich Physik, Institut f{\"u}r Kernphysik, 64289 Darmstadt, Germany}

\affiliation[3]{IPN Orsay, 15 rue Georges Clemenceau, 91406, Orsay, France}

\affiliation[4]{CEA Saclay, IRFU/DPhN, Centre de Saclay, 91191, Gif-sur-Yvette, France}
\affiliation[5]{CINTECX, Universidade de Vigo, E-36200 Vigo, Spain}
\affiliation[6]{CEA DAM,DIF, F-91297 Arpajon, France}
\affiliation[7]{GSI Helmholtzzentrum f{\"u}r Schwerionenforschung, Planckstra\ss e 1, 64291, Darmstadt, Germany}
\affiliation[8]{Lund University, Lund, Sweden}
\affiliation[9]{LIP and Faculty os Sciences, University of Lisbon, 1000-149 Lisbon, Portugal}
\affiliation[10]{RBI Zagreb, Bijenicka cesta 54, HR10000, Zagreb, Croatia}
\affiliation[11]{Technische Universit{\"a}t M{\"u}nchen, James-Franck-Str 1, 85748, Garching, Germany}
\affiliation[12]{Institutionen f{\"o}r Fysik, Chalmers Tekniska H{\"o}gskola, 412 96 G{\"o}teborg, Sweden}
\affiliation[13]{University of Guelph, 50 Stone Road E, N1G 2W1, Guelph, ON, Canada}
\affiliation[14]{Institut für Kernphysik der Universität zu Köln, Zülpicher Strasse 77, D-50937 Köln, Germany}
\affiliation[15]{School of Physics, Engineering and Technology, University of York, YO10 5DD York, UK}
\affiliation[16]{Institute for Basic Science, Center for Exotic Nuclear Studies, 34126, Daejeon, Korea (Republic of)}
\affiliation[17]{Instituto de Estructura de la Materia, CSIC, E-28006 Madrid, Spain}
\affiliation[18]{INFN Laboratori Nazionali del Sud, Via Santa Sofia 62, 95123, Catania, Italy}
\affiliation[19]{Jozef Stefan Institute, Jamova cesta 39,
1000 Ljubljana, Slovenia}

\emailAdd{antia.grana.gonzalez@usc.es}

\FullConference{%
  FAIR next generation scientists - 7th Edition Workshop (FAIRness2022)\\
  23-27 May 2022\\
  Paralia (Pieria, Greece)
}

\begin{document}
\maketitle

\paragraph{}
Despite the recent experimental and theoretical progress in the investigation of the nuclear fission process, a complete description still represents a challenge in nuclear physics because it is a very complex dynamical process, whose description involves the coupling between intrinsic and collective degrees of freedom, as well as different quantum-mechanical phenomena. To improve on the existing data on nuclear fission, we produce fission reactions of heavy nuclei in inverse kinematics by using quasi-free (p,2p) scattering, which induce fission through particle-hole excitations that can range from few to ten's of MeV. The measurement of the four-momenta of the two outgoing protons allows to reconstruct the excitation energy of the fissioning compound nucleus and therefore to study the evolution of the fission yields with temperature. The realization of this kind of experiment requires a complex experimental setup, providing full isotopic identification of both fission fragments and an accurate measurement of the momenta of the two outgoing protons. This was realized recently at the GSI/FAIR facility and here some preliminary results are presented.
\section{Introduction}

Nuclear fission plays an important role in the nucleosynthesis r-process, which was indicated as the mechanism responsible for the production of the heaviest elements in the universe. Therefore, fission studies in the neutron-rich region are required in order to measure fission yields and fission barrier heights to improve r-process calculations \cite{r-process, r-process-Pinedo}. \\
In the last decade, unprecedented fission experiments have been carried out at the GSI/FAIR facility using the inverse-kinematics technique in combination with state-of-the-art detectors. For the first time in the history of fission, it was possible to simultaneously measure and identify both fission fragments in terms of their mass and atomic numbers \cite{JoseLuis2015, Pellereau} and to extract correlations between fission observables sensitive to the dynamics of the fission process \cite{JoseLuis2016} and the nuclear structure at the scission point  \cite{Audrey, Audrey2020}. However, a limitation of those previous experiments was the impossibility of measuring the excitation energy of the compound nucleus undergoing fission, an observable that plays a fundamental role in the description of the process. Depending on the amount of available energy, different regions of the potential landscape can be populated, leading to different fission paths and consequently to different fission yields distributions. The gradual weakening of the influence of shell structure on the fission yields with the increasing of excitation energies can be quantified with the so-called suppression function \cite{sup}, which suppresses the microscopic term of the energy-dependent effective potential. However, the model parameters used in such an approach are still not constrained due to the lack of data at different excitation energies. This lack of data is caused by the use of Coulomb excitation to induce fission of exotic nuclei and the use of transfer- and fusion-fission reactions only with stable nuclei. We went further by combining fission with quasi-free (p,2p) reactions, which allows for the first time to measure many excitation energies with stable and exotic nuclei providing therefore very valuable data to constrain the suppression function parameters. Another interesting question is the possible dependence of the energy sorting of the fission fragments on the excitation energies of the compound nucleus, since in the studies in direct kinematics with neutrons, the well known saw-tooth shape shows higher neutron multiplicities for higher incident neutron energies \cite{saw_tooth}.

In this work, the main scientific goal is to investigate the energy dependence of the fission yields. Such data can be obtained by using quasi-free (p,2p) reactions and combining the previous experimental setup with a silicon tracker based on AMS-type detectors \cite{Alcaraz} and the calorimeter CALIFA (CALorimeter for In-Flight detection of gamma-rays and high energy charged pArticles) \cite{Califa} developed by the R³B collaboration. Here we report on the analysis of the first fission experiment performed at the GSI/FAIR accelerator facilities in Darmstadt (Germany) using the quasi-free (p,2p) reactions as mechanism to induce fission of $^{238}$U projectiles.

\section{Experiment}

The experiment was performed in March 2021 by using primary beams of $^{238}$U at 560$A$ MeV delivered by the SIS18 synchrotron, which were guided to the experimental area to impinge on a liquid hydrogen target in order to produce quasi-free (p,2p) reactions. We use the inverse-kinematics technique in which the heavy nucleus undergoing fission is the projectile. This methodology was developed at GSI and represents a major breakthrough in the characterization of the fission fragments. The major advantage is that the fission fragments are emitted in a narrow cone in the forward direction with very high velocities, permitting to detect both fragments simultaneously and to measure the isotopic yields.\\

\begin{center}
\begin{figure}[h]
\begin{center}

\includegraphics[width=0.80\textwidth,keepaspectratio]{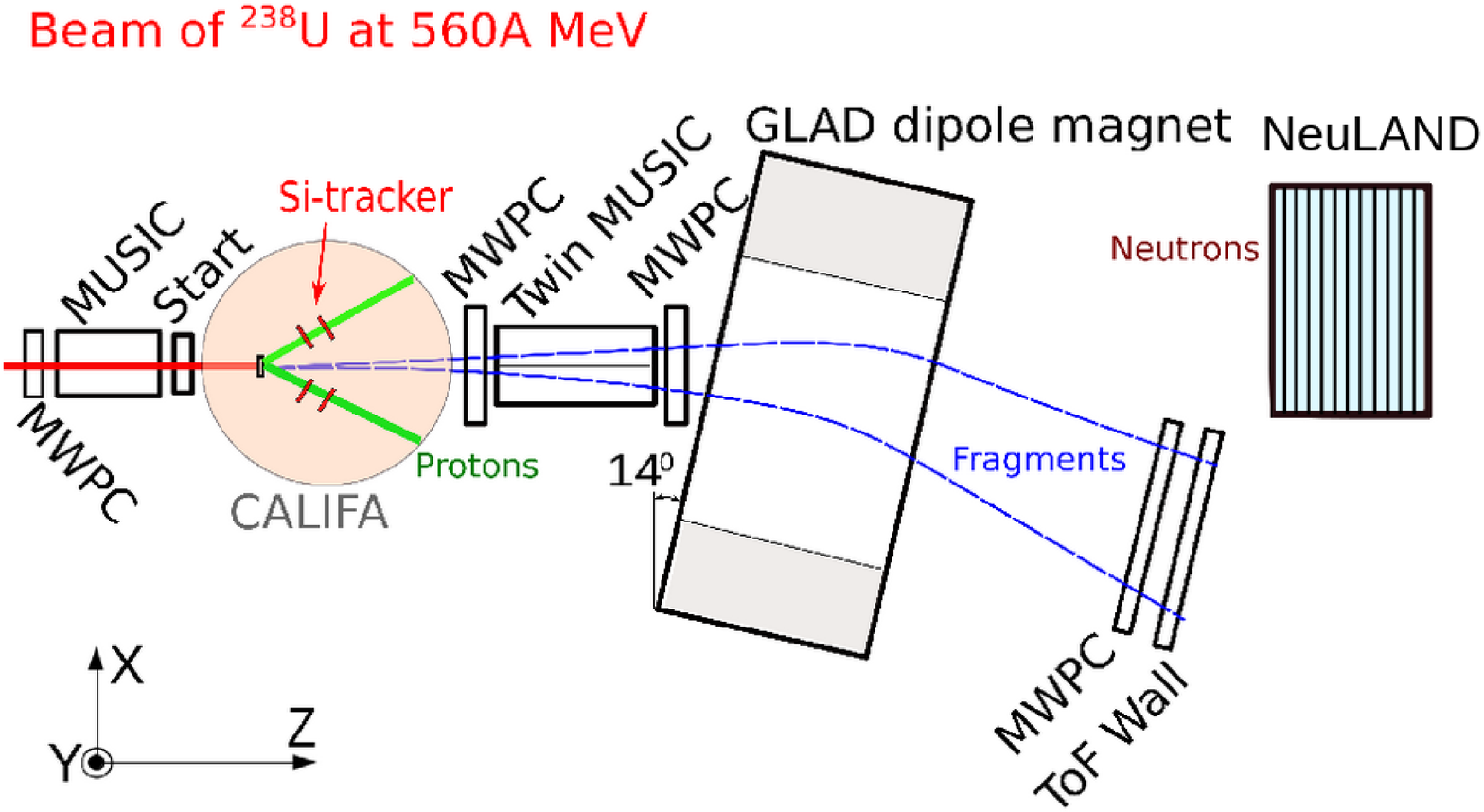}
\caption{Schematic view of the experimental setup used in the present work to study quasi-free (p,2p) induced fission reactions.
}
\label{setup}
\end{center}
\end{figure}
\end{center}

Using this methodology, previous fission experiments at GSI already provided for the first time the complete isotopic identification of both fission fragments.
In the (p,2p)-fission experiment the fragments are identified by using the same detectors. The experimental setup (see the Fig \ref{setup}) consists of several multi-wire proportional chambers (MWPCs) to track the trajectories before and after the superconducting dipole GLAD, a Twin-MUSIC (Multi-Sampling Ionization Chamber) divided into two sections which are segmented into 16 anodes to measure the charges of the fission fragments and a ToF-Wall composed by 28 plastic scintillators to measure the time of flight. With all these measurements, employing the $\Delta E-B\rho-ToF$ technique, the mass of the fission fragments can be obtained, allowing therefore for full isotopic identification. Neutrons emitted from the fission fragments are detected in NeuLAND (New Large-Area Neutron Detector) \cite{neuland}, located 15m downstream from the target.\\
Since our goal is to study the energy dependence of the fission yields, we access the excitation energy via the aforementioned quasi-free (p, 2p) reactions which are knockout reactions between protons of the liquid hydrogen target and protons bound in the heavy projectile nucleus. The products of the reaction are the two outgoing protons and the excited compound nucleus characterized by the particle-hole excitations created by the removed proton. This excited compound nucleus can then de-excite via fission or neutron evaporation. From the measurement of the momenta of the two outgoing protons the excitation energy of the compound nucleus can be reconstructed using the missing energy method \cite{Missing_energy}.
In order to obtain the four-momenta, the energy and the tracking of the protons are required. To reconstruct the protons trajectory a silicon tracker is located in front of the target, which was equipped with two double-plane arms consisting of an array of three AMS-type \cite{Alcaraz} 0.3 mm thick double-sided silicon-strip detectors. Then to measure the energy of the protons the CALIFA calorimeter \cite{Califa} is surrounding the whole, which consists of 1504 CsI crystal scintillators covering a polar angle range between 22 and 90 degrees.

\section{Results}

\begin{center}
\begin{figure}[b!]
\begin{center}
\includegraphics[width=0.49\textwidth,keepaspectratio]{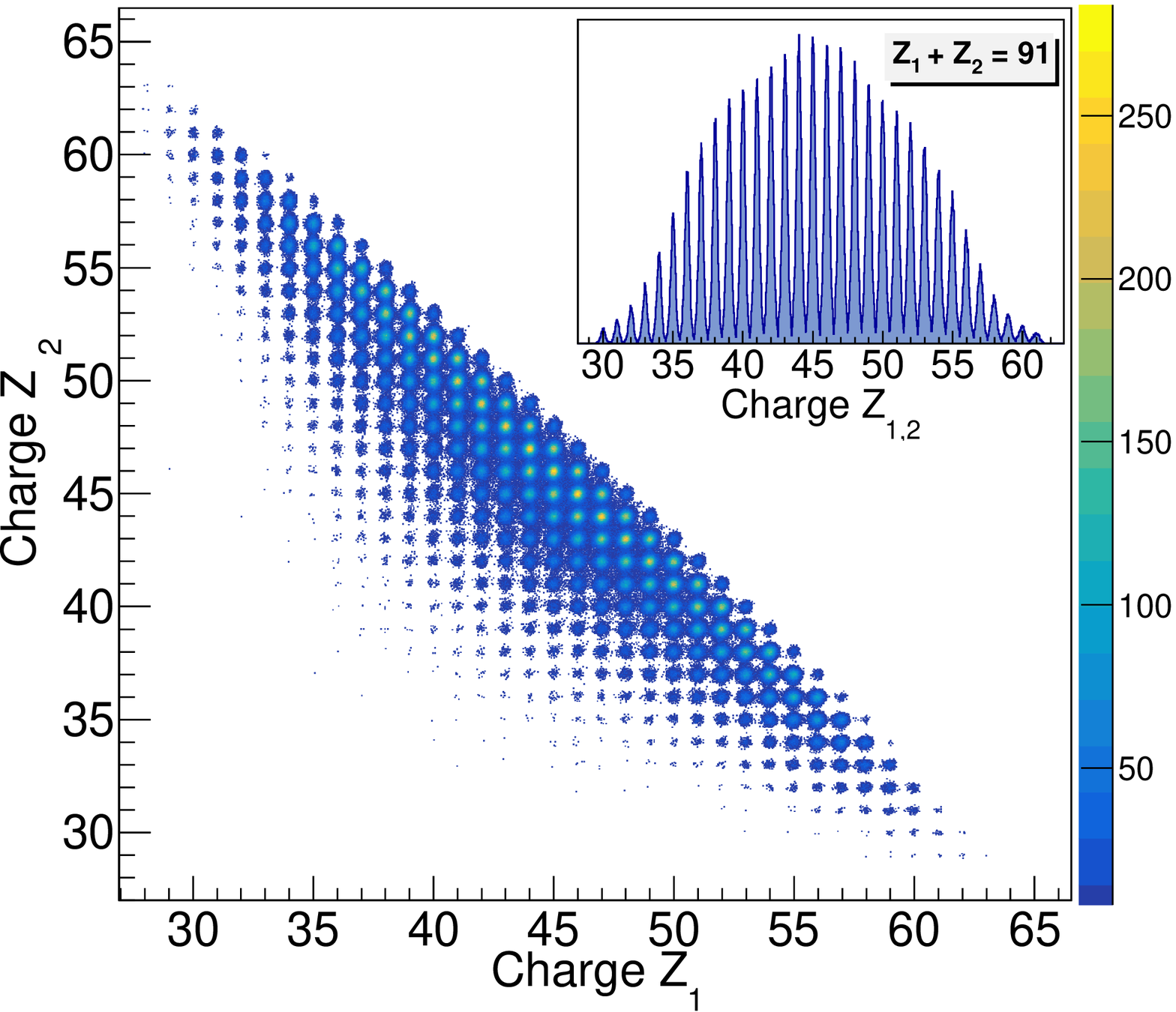} 
\includegraphics[width=0.49\textwidth,keepaspectratio]{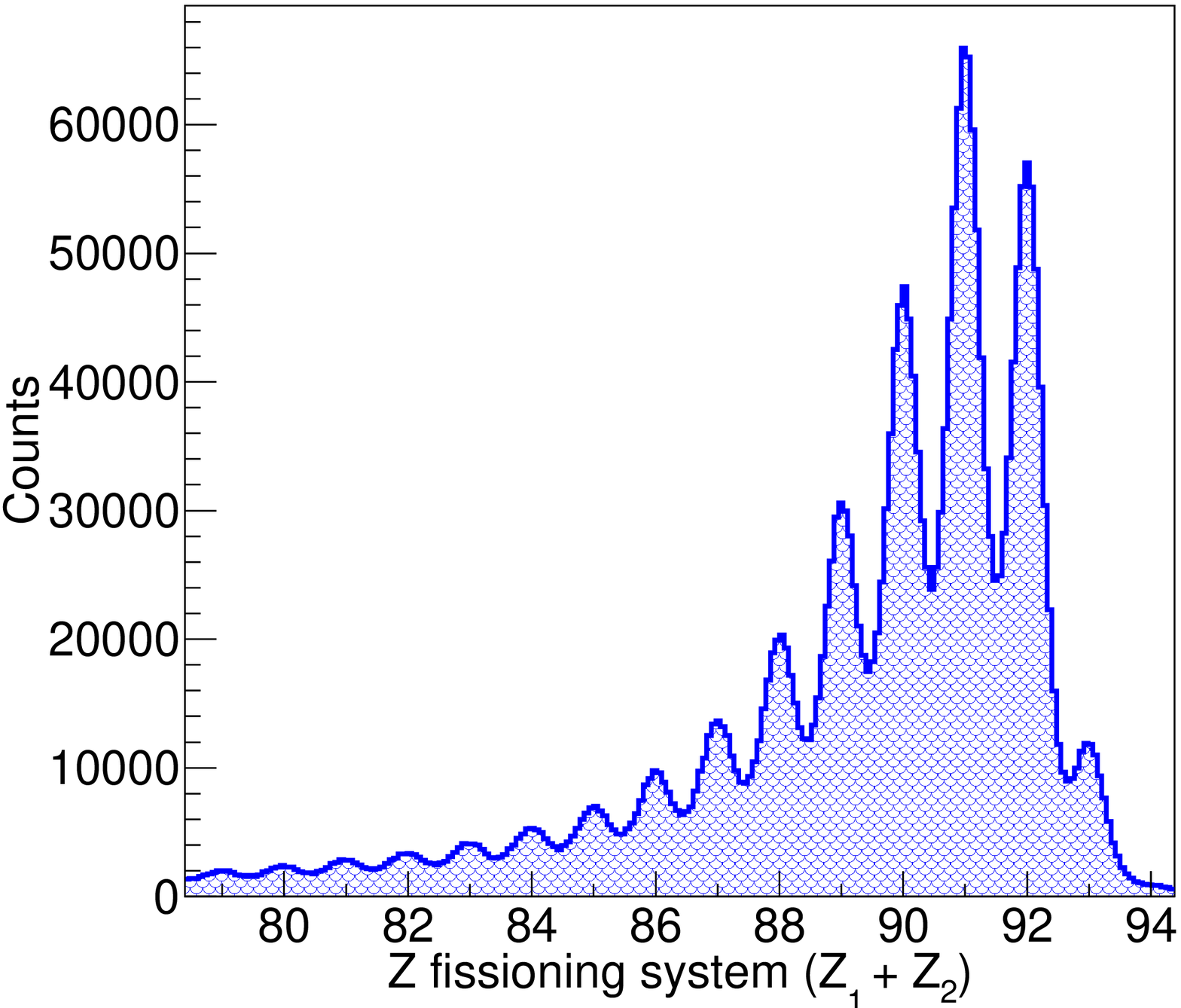} 
\caption{Left: Two-dimensional cluster plot of the charges $Z_1$ and $Z_2$ of both fission fragments for all fissioning systems. The inset shows the fission fragments distribution for $Z_{1} + Z_{2} = 91$. Right: Fissioning systems distribution.
}
\label{Zff}
\end{center}
\end{figure}
\end{center}

The first step of the analysis is to obtain the charge identification of both fission fragments given by the Twin-MUSIC, which is designed with two independent ionization chambers divided into two sections in order to measure the atomic number of both fission fragments simultaneously. The correlation of the derived charge for both fission fragments $Z_1$ and $Z_2$ is shown in the Fig.~\ref{Zff}-left in a two-dimensional cluster plot. Each diagonal of clusters in this figure corresponds to a different fissioning system, and the fissioning system distribution is displayed in the Fig.~\ref{Zff}-right. Since the beam is $^{238}_{92}$U, when the desired (p,2p) takes place the resulting nucleus is an excited $^{237}_{91}$Pa, which is the $Z_{1} + Z_{2} = 91$ peak, the one with the highest statistics of this plot. Nevertheless, it is important to notice that this peak might not have only $^{237}_{91}$Pa systems, it will have the contribution for other lighter $_{91}$Pa isotopes as well, since the $^{237}_{91}$Pa can also de-excite via neutron emission and undergo fission after. It is also possible to obtain different excited $_{91}$Pa isotopes when the reaction is not quasi-free. A selection in the opening angle between the protons serves to select the quasi-free reactions in which there is no evaporation of neutrons before fission.\\
The inset of the Fig.~\ref{Zff}-left shows the fission fragment distribution for the production of $_{91}$Pa, with a charge resolution of 0.38 units (FWHM). The distribution fits well to an overall Gaussian shape, although two small humps around Z=40 and Z=50 are slightly noticeable. These two peaks are the asymmetric fission contribution and they appear because of the shell effects \cite{schmidt} which favour the formation of nuclei with magic numbers around Z=50 in this case. Without any constrain on the energy (like in this plot) the symmetric contribution from the high energies will partially shadow the asymmetric humps and the total distribution will be the convolution of the three Gaussian distributions. Nevertheless, when this plot is constrained to low energies, the symmetric component should gradually vanish, since the fission barrier in $_{91}Pa$ is supposed to be higher for symmetric fission. Therefore, the asymmetric distribution would start to be more and more noticeable and at some point it would be the only visible one. Once the excitation energy is reconstructed, it will be possible to plot this same graph for different excitation energies and to study the evolution of the fission yields from asymmetric to symmetric while going from low to high energies.\\

%
\section{Conclusions}

The first (p,2p)-fission experiment has been carried out at GSI/FAIR with projectiles of $^{238}$U at energies of $560A$ MeV impinging on a liquid hydrogen target and using the R³B experimental setup to measure all the reaction products. These complete kinematics measurements will allow to correlate the fission yields with the excitation energy of the fissioning compound nucleus and to study the energy sharing between the two fission fragments. Additionally, this experiment permits to develop the data analysis and validate the methodology to use this approach in future experiments with exotic nuclei.
The current state of the analysis is that the charge identification is completed and the next steps will be the identification in mass number and the reconstruction of the excitation energy spectrum. In the near future further (p,2p)-fission experiments are planned to be carried out at GSI/FAIR with exotic neutron-rich projectiles close to the neutron shell N=152 to obtain their fission yields and fission barrier heights for constraining r-process calculations.

\section{Acknowledgments}
This work was partially supported by the Spanish Ministry for Science and Innovation under grants PGC2018-099746-B-C21 and PGC2018-099746-B-C22, by the Regional Government of Galicia under the program “Grupos de Referencia Competitiva” ED431C-2021-38 and by the “María de Maeztu” Units of Excellence program MDM-2016-0692. J.L.R.S. thanks the support from Xunta de Galicia under the program of postdoctoral fellowship ED481D-2021-018.

\end{document}